\documentclass[12pt]{article}

\usepackage{pstricks}
\usepackage{cancel}
\usepackage{amssymb,amsmath}

\def\harr#1#2{\smash{\mathop{\hbox to .3in{\rightarrowfill}}
 \limits^{\scriptstyle#1}_{\scriptstyle#2}}}

\def\s2{\frac{1}{\sqrt2}}

\def\be{\begin{equation}}
\def\ee{\end{equation}}
\def\beqa{\begin{eqnarray}}
\def\eeqa{\end{eqnarray}}

\def\Dsl{\,\raise.15ex\hbox{/}\mkern-13.5mu D} 
\def\d3{d^3}

\begin{document}

\vspace{.5cm}
\begin{center}
\Large{\bf  Ground-state Wigner functional of linearized gravitational field}\\
\vspace{1cm}

\large Hugo Garc\'{\i}a-Compe\'an$^a$\footnote{e-mail address: {\tt
compean@fis.cinvestav.mx}}, Francisco J. Turrubiates$^b$\footnote{e-mail address: {\tt
fturrub@esfm.ipn.mx}}
\\
[2mm]
{\small \em $^a$Departamento de F\'{\i}sica, Centro de
Investigaci\'on y de Estudios Avanzados del IPN}\\
{\small\em P.O. Box 14-740, 07000 M\'exico D.F., M\'exico.}
\\[4mm]
{\small \em $^b$Departamento de F\'{\i}sica, Escuela Superior de F\'{\i}sica
y Matem\'aticas del IPN}\\
{\small\em Unidad Adolfo
L\'opez Mateos, Edificio 9, 07738, M\'exico D.F., M\'exico.}
\\[4mm]
\vspace*{2cm}
\small{\bf Abstract} \\
\end{center}
The deformation quantization formalism is applied to the linearized
gravitational field. Standard aspects of this formalism are worked
out before the ground state Wigner functional is obtained. Finally,
the propagator for the graviton is also discussed within the context
of this formalism.

\begin{center}
\begin{minipage}[h]{14.0cm} { }
\end{minipage}
\end{center}

\bigskip
\bigskip

\vspace{3cm}

\leftline{September 5, 2011}

\newpage

\section{Introduction}
The study of weak gravitational fields has produced important
results during the years. One of them is the prediction of
gravitational waves for the non-stationary cases. This has been a
very important area of research and many groups around the world are
working on its detection through bars and interferometers  (for a
review involving the the experimental and theoretical parts, see for
instance, \cite{Misner:1974qy,Will:2006,Schutz:2009}). It is
expected that they will be a reality within the next few years.

From the theoretical point of view  ADM's formalism
\cite{Arnowitt:1962hi,Deser:1959zza} allowed for the first time to
deal the gravitational field from a suitable point of view to study
its mathematical structure. In particular the weak field was also
incorporated into this framework. In the present case we are not
considering the study of sources by themselves and all the analysis
is performed in the vacuum. Many questions can also be treated for
this case just as finding the corresponding ground state of the
field, that is the vacuum. Some remarkable results dealing with this
problem had been obtained previously by Kuchar \cite{Kuchar:1970mu}
and Hartle \cite{Hartle:1984ke}.  In these works the canonical
quantization and path integral procedures were employed to obtain
the ground state wave function of linearized gravity. However there
exists another method of quantization the so called deformation
quantization formalism \cite{bffls} that can be also applied to the
gravitational field \cite{antonsenuno,antonsendos}. In a recent work
we employed such technique  to deal with some cosmological models in
the minisuperspace \cite{Cordero:2011xa}. The present paper is
motivated by the possibility to extend some of the results that we
obtained there to some other interesting cases in general relativity
and its approaches.

The ideas of deformation quantization are based on the fundamental
works of Weyl, Wigner, Groenewald, Moyal, Gerstenhaber and Vey (for
some recent reviews and developments on the subject, see for
instance, \cite{Dito:2002dr,bookzfc}) that evolved during the years
until the introduction of the theory in its final form given by
Bayen \textit{et al} \cite{bffls}. This quantization comes from a
deformation of the usual algebra product of smooth functions on the
classical phase space which then induce a deformation of the Poisson
bracket algebra. It has been proved that this new product called the
$\star-$product exists for any symplectic or Poisson manifold
\cite{f,k}. In terms of this quantization it is not necessary the
use of operators or even the introduction a Hilbert space, which in
principle offers important advantages when we deal with systems
whose phase space has non-trivial topology \cite{f,k}. For these
reasons we consider that this method is suitable to deal with
certain problems at quantum level of the gravitational field in the
phase space.

The main goal of the present paper is to obtain by this procedure
the ground state Wigner functional of linearized gravity and compare
it with the previous results \cite{Kuchar:1970mu,Hartle:1984ke}. At
the same time we look for to present a new example of the
application of the deformation quantization formalism, in particular
to field theory, since there are just few of them. Finally, we also
hope that the results presented here will motivate the application
of this method of quantization to different configurations and
problems involving the gravitational field.

Our paper is organized as follows. In section 2 we introduce the
conventions and notation of the linearized gravitational field
required to apply the deformation quantization formalism. The
deformation quantization of the linearized gravitational field is
performed in the temporal gauge and in the $(TT)$-gauge in section
3. In particular, in this same section, the Wigner functional for
the ground state, the normal ordering operator and the graviton
propagator are obtained. Finally in section 4 we give our
conclusions and final remarks.


\section{Overview of the Linearized Gravitational Field}

We start by considering the Fierz-Pauli action defined in a
Minkowski spacetime with signature $(-,+,+,+)$
\begin{equation}
{\cal L} = -{1 \over 4} h_{\alpha \beta, \lambda} h^{\alpha \beta}_{ \ \ , \lambda}
-{1 \over 2} h^{\alpha}_{ \ \alpha, \lambda} h^{\lambda \beta}_{ \ \  \ , \beta}
+ {1 \over 4} h^{\alpha}_{ \  \alpha , \lambda} h_{ \ \beta}^{ \beta \ , \lambda} +
{1 \over 2} h_{ \  \beta , \alpha}^{\alpha} h^{\beta \lambda}_{ \ \ , \lambda},
\label{fpaction}
\end{equation}
where $\alpha, \beta,$ etc. $=0,1,2,3$ and $i,j,$ etc.$=1,2,3$.
Along the paper it is assumed the Einstein's convention on the sum
over the repeated indices. Moreover we have used the notation
$h_{\alpha \beta, \lambda}= \nabla_\lambda h_{\alpha \beta}$. The
fields $h_{\alpha \beta}(\vec{x},t)$ are dynamical, in particular
for a $(3+1)$-decomposition of spacetime, $h_{00}$, $h_{0i}$ and
$h_{ij}$ are the relevant variables. The canonical momenta
corresponding to these fields are
\begin{equation}
\pi_{00}={\delta L \over \delta \dot{h}_{00}}, \ \ \ \ \
\pi_{0i}={\delta L \over \delta \dot{h}_{0i}}, \ \ \ \ \
\pi_{ij}={\delta L \over \delta \dot{h}_{ij}},
\end{equation}
where $\dot{h}_{ij}=h_{ij,0}={\partial h_{ij} \over \partial x^0}$
is the trace of $h_{ij}$.

They can be computed from the action (\ref{fpaction}) and they take
the following explicit form
\begin{equation}
\pi_{00}={1\over 2} h_{0k,k}, \ \ \ \ \  \pi_{0i}={1\over 2}
h_{ll,k} - {1\over 2} h_{00,k} - h_{lk,l}, \ \ \ \ \  \pi_{ij}=
{1\over 2} \big[ \dot{h}_{ij} + \delta_{ij}(-\dot{h}_{kk}
+h_{k,k})\big],
\label{tresecs}
\end{equation}
where $h_{kk}={\rm Tr}(h_{ij}).$

The corresponding Hamiltonian  can be written as
\begin{eqnarray}
H&=&\int d^3x {\cal H} \nonumber \\
&=& \int d^3 x \big( \pi_{00} \dot{h}_{00} + \pi_{0k} \dot{h}_{0k} +
\pi_{ij} \dot{h}_{ij}\big) - L,
\label{ham}
\end{eqnarray}
where $L=\int d^4x{\cal L}$ and the hamiltonian density reads
$$
{\cal H}= \pi_{ij} \pi_{ij} - {1 \over 2} \pi_{mm} \pi_{nn} + {1
\over 2} \pi_{mm} h_{0k,k} + {1 \over 8} h_{0k,k} h_{0l,l} + {1
\over 2} h_{00,m} h_{ll,m} - {1 \over 4} h_{ll,m} h_{kk,m}
$$
\begin{equation}
- {1 \over 2} h_{0k,m} h_{0k,m} + {1 \over 4} h_{kl,m} h_{kl,m} - {1
\over 2} h_{ml,m} h_{nl,n} - {1 \over 2} h_{00,n} h_{mn,m} +  {1
\over 2} h_{mn,m} h_{ll,n}.
\label{hamdensity}
\end{equation}

In the present paper we are going to work in the {\it temporal
gauge} where $h_{\alpha 0}=0$. In this gauge we have that the
Poisson bracket (at the same fixed time) between $h_{ij}(\vec{x},t)$
and $\pi^{kl}(\vec{x},t)$ is given by
\begin{equation}
\{h_{ij}(\vec{x},t), \pi^{kl}(\vec{x}',t) \}_{PB}= \delta_{ij}^{ \ \
kl}\delta(\vec{x}-\vec{x}')\,\, ,
\label{pbone}
\end{equation}
where $\delta_{ij}^{ \ \ kl}= \frac{1}{2} (\delta^k_i \delta^l_j +
\delta^k_j \delta^l_i)$. The rest of Poisson brackets vanish
\begin{equation}
\{h_{ij}(\vec{x},t), h_{kl}(\vec{x}',t) \}_{PB} = 0 =
\{\pi^{ij}(\vec{x},t), \pi^{kl}(\vec{x}',t) \}_{PB}. \label{pbtwo}
\end{equation}
The standard expansion of the field variables in the temporal gauge reads
\begin{equation}
h_{ij}(\vec{x},t) =  \int {d^3k \over (2 \pi)^{3/2}} \bigg({ \hbar
\over 2 \omega(\vec{k})}\bigg)^{1/2}  \bigg(
\mathfrak{h}_{ij}(\vec{k},t) \exp\big(i\vec{k}\cdot \vec{x} \big) +
\mathfrak{h}^*_{ij}(\vec{k},t) \exp \big(-i\vec{k} \cdot \vec{x}
\big) \bigg),
\end{equation}
where $\omega(\vec{k}) = |\vec{k}|$, $\mathfrak{h}_{ij}(\vec{k},t) =
\mathfrak{h}_{ij}(\vec{k}) \exp \big\{ -i \omega(\vec{k}) t \big\}$
and  $\vec{k} \cdot \vec{x} \equiv k_jx_j$. The momentum is given by

\begin{equation}
\pi^{ij}(\vec{x},t) = \int {d^3k \over (2 \pi)^{3/2}} \sqrt{\hbar
\over 2 \omega(\vec{k})} \bigg(
\mathfrak{p}^{ij}(\vec{k},t)\exp\big(i\vec{k}\cdot \vec{x} \big) +
\mathfrak{p}^{*ij}(\vec{k},t) \exp \big(-i\vec{k} \cdot \vec{x}
\big) \bigg), \label{momento}
\end{equation}
where $\mathfrak{p}^{ij}(\vec{k},t)= \mathfrak{p}^{ij}(\vec{k}) \exp \big\{ -i \omega(\vec{k}) t
\big\}$. From Eq. (\ref{tresecs}) one can find
\begin{equation}
\mathfrak{p}_{ij}(\vec{k},t)= -{i \omega \over 2} \mathfrak{h}_{ij}(\vec{k},t) + {i \omega \over 4} {\delta}_{ij} \mathfrak{h}_{kk}(\vec{k},t) + {i \over 2} \big[\delta_{ij} k_k \mathfrak{h}_{0k}(\vec{k},t) - \big(k_j \mathfrak{h}_{0i}(\vec{k},t) + k_i \mathfrak{h}_{0j}(\vec{k},t)\big)\big].
\label{momentok}
\end{equation}

We will work with the decomposition into the transverse-traceless
part ($TT$), the transverse part ($T$) and the longitudinal part
($L$) of this field
\begin{equation}
h_{ij}(\vec{x},t) = h_{ij}^{(TT)}(\vec{x},t) +
h_{ij}^{(T)}(\vec{x},t) + h_{ij}^{(L)}(\vec{x},t).
\label{separation}
\end{equation}
Here the $(TT)$ part satisfies the following conditions:
\begin{equation}
h^{(TT)}_{\alpha 0}=0, \  \ \ \ \ h^{i \ (TT)}_{ \ i} =0,  \ \ \  \
\     h_{ij,k}^{(TT)} =0,
\label{normaTT}
\end{equation}
which is called the $(TT)$-gauge.

It is well known that in order to work out with this $(TT)$-gauge is
more appropriate (but equivalent) to solve the Gauss law constraint
\begin{equation}
\nabla_j \pi^{ij}(\vec{x})=0.
\label{gausslaw}
\end{equation}

Now the decomposition (\ref{separation}) can be introduced through
the introduction of a linear projection operator
\begin{equation}
P_{jk}= \delta_{jk} - {k_j k_k \over |\vec{k}|^2},
\end{equation}
with the following properties:
\begin{equation}
P_{jl} P_{lk}= P_{jk}, \ \ \ \   P_{jl}k_l=0, \ \ \ \ P_{jj}=2.
\end{equation}
The transverse traceless ($TT$), transverse ($T$) and longitudinal
($L$) parts can be written in terms of the projection operator as
$$
\mathfrak{h}_{jk}^{(TT)}(\vec{k},t)= P_{jl}P_{mk}\mathfrak{h}_{lm}(\vec{k},t) -{1\over 2}P_{jk}(P_{lm}\mathfrak{h}_{lm}(\vec{k},t)),
$$
$$
\mathfrak{h}^{(T)}_{jk}(\vec{k},t)={1\over 2} P_{jk}(P_{lm}\mathfrak{h}_{lm}(\vec{k},t)),
$$
\begin{equation}
\mathfrak{h}^{(L)}_{jk}(\vec{k},t) = \mathfrak{h}_{jk}(\vec{k},t) - P_{jl}P_{mk} \mathfrak{h}_{lm}(\vec{k},t).
\end{equation}

If one fixes the ($TT$) gauge, only the ($TT$) components of
$h_{ij}$ survive and it results
\begin{equation}
h^{(TT)}_{ij}(\vec{x},t) =  \int { d^3k\over (2 \pi)^{3/2}} \bigg({
\hbar \over 2 \omega(\vec{k})}\bigg)^{1/2}  \bigg(
\mathfrak{h}^{(TT)}_{ij}(\vec{k},t) \exp\big(i\vec{k}\cdot \vec{x}
\big) + \mathfrak{h}^{*(TT)}_{ij}(\vec{k},t) \exp \big(-i\vec{k}
\cdot \vec{x} \big) \bigg),
\end{equation}
and
$$
\pi^{(TT)}_{ij}(\vec{x},t) = {1 \over 2} \dot{h}^{(TT)}_{ij}(\vec{x},t)
$$
\begin{equation}
= \int { d^3k\over (2 \pi)^{3/2}} \bigg({i \over 2} \bigg)\bigg({
\hbar \omega(\vec{k}) \over 2}\bigg)^{1/2} \bigg( -
\mathfrak{h}^{(TT)}_{ij}(\vec{k},t) \exp\big(i\vec{k}\cdot \vec{x}
\big) + \mathfrak{h}^{*(TT)}_{ij}(\vec{k},t) \exp \big(-i\vec{k}
\cdot \vec{x} \big) \bigg).
\end{equation}
In this $(TT)$-gauge the Poisson brackets (\ref{pbone}) should be
severely modified since they must satisfy the Gauss' law
(\ref{gausslaw}). Then
\begin{equation}
\{h^{(TT)}_{ij}(\vec{x}), \pi^{kl(TT)}(\vec{x}') \}_{PB}= \delta^{ \
\ kl(TT)}_{ij}\delta(\vec{x}-\vec{x}')\,\, ,
\label{poissonTT}
\end{equation}
where $\delta^{ \ \ kl(TT)}_{ij}$ is the transverse-traceless Dirac
function and it is given by
\small
\begin{equation}
\delta^{ \ \ kl(TT)}_{ij}= \frac{1}{2} \bigg[\bigg(\delta^k_i -
{\partial^k \partial_i \over \nabla^2}\bigg) \bigg(\delta^l_j -
{\partial^l \partial_j \over \nabla^2}\bigg) + \bigg(\delta^k_j -
{\partial^k \partial_j \over \nabla^2}\bigg) \bigg(\delta^l_i -
{\partial^l \partial_i \over \nabla^2}\bigg) -  \bigg(\delta_{ij} -
{\partial_i \partial_j \over \nabla^2}\bigg) \bigg(\delta^{kl} -
{\partial^k\partial^l \over \nabla^2}\bigg)\bigg].
\label{deltaTT}
\end{equation}
\normalsize

Then working in the temporal gauge the field $h_{ij}(\vec{x},t)$ can
be written as
$$
h_{ij}(\vec{x},t) =  \int {d^3k \over (2 \pi)^{3/2}} \bigg\{\
\varepsilon^{ab}_{ij} \bigg(
 \mathfrak{h}^{(TT)}_{ab}(\vec{k},t)\exp\big(i \vec{k}\cdot \vec{x} \big) +
  \mathfrak{h}^{*(TT)}_{ab}(\vec{k},t)\exp\big(- i \vec{k}\cdot \vec{x} \big)\bigg)
$$
\begin{equation}
+ \bigg(\mathfrak{h}^{(T)}_{ij}(\vec{k},t)+  \mathfrak{h}^{*(T)}_{ij}(-\vec{k},t) + \mathfrak{h}^{(L)}_{ij}(\vec{k},t)+  \mathfrak{h}^{*(L)}_{ij}(-\vec{k},t) \bigg) \exp\big(i \vec{k}\cdot \vec{x} \big) \bigg\},
\end{equation}
where $\varepsilon^{ab}_{ij}= e^{a}_i(\vec{k})e^{b}_j(\vec{k})$ is
symmetric and traceless in $a$ and $b$ with $a,b=1,2$. Here
$e^a_i(\vec{k})$\footnote{The polarization vectors $e^a_i(\vec{k})$
correspond to ${\bf e}_+$ and ${\bf e}_\times$ for $a=1$ and $a=2$
respectively.} are the two well known polarization vectors expanded
by the $(TT)$-part. These vectors are orthogonal
\begin{equation}
e^{a}_i(\vec{k}) e^{b}_i(\vec{k}) = \delta_{ab}, \ \ \ \ k_i
e^{a}_i(k) = 0.
\end{equation}

Now the Hamiltonian (\ref{ham}) in terms of the decomposition
(\ref{separation}) reads
\begin{equation}
H=\int d^{3}x
\big(\pi^{(TT)}_{ij}\pi^{(TT)}_{ij}+\frac{1}{4}h^{(TT)}_{kl,m}h^{(TT)}_{kl,m}\big)+\int
d^{3}x \mathcal{H}_{LG},
\label{hamiltoniano}
\end{equation}
where
\small
\begin{eqnarray}
&&\int d^{3}x \mathcal{H}_{LG}= \int d^{3}k \frac{\hbar}{8 \omega(\vec{k})} \left\{k_{k}\frac{\omega}{2}\mathcal{B}_{mm}^{*}(\vec{k},t)\mathcal{A}_{0k}(\vec{k},t)+ k_{m}k_{k}\mathcal{A}_{0k}(\vec{k},t)\mathcal{A}_{0m}^{*}(\vec{k},t) \right. \nonumber  \\
&&\left. -\frac{1}{2} \left[\frac{\omega}{2}(\mathcal{C}_{mm}(\vec{k},t)\mathcal{B}_{nn}^{*}(\vec{k},t) +\mathcal{C}_{mm}^{*}(\vec{k},t)\mathcal{B}_{nn}(\vec{k},t))+ k_{n}(\mathcal{C}_{mm}(\vec{k},t)\mathcal{A}_{0n}^{*}(\vec{k},t)
+\mathcal{C}_{mm}^{*}(\vec{k},t)\mathcal{A}_{0n}(\vec{k},t)) \right] \right\} \nonumber \\
&& +\int d^{3}k \frac{\hbar}{4\omega(\vec{k})} \left\{\frac{k_{k}k_{l}}{4}\mathcal{A}_{0k}^{*}(\vec{k},t)\mathcal{A}_{0l}(\vec{k},t)+
|\vec{k}|^{2}\mathcal{A}_{00}^{*}(\vec{k},t)\mathcal{A}_{ii}(\vec{k},t)
- \frac{1}{2} |\vec{k}|^{2}\mathcal{A}_{ii}^{*}(\vec{k},t)\mathcal{A}_{jj}(\vec{k},t) \right. \nonumber \\
&& \left. -
\mathcal{A}_{0k}^{*}(\vec{k},t)\mathcal{A}_{0k}(\vec{k},t) -
k_{m}k_{n}(\mathcal{A}_{ml}^{*}(\vec{k},t)\mathcal{A}_{nl}(\vec{k},t)
+ \mathcal{A}_{00}^{*}(\vec{k},t)\mathcal{A}_{mn}(\vec{k},t)-
\mathcal{A}_{mn}^{*}(\vec{k},t)\mathcal{A}_{ii}(\vec{k},t))
\right\}.
\label{densidadham}
\end{eqnarray}
\normalsize Here the coefficients can be expressed by
\small
\begin{eqnarray}
&&\mathcal{A}_{00}(\vec{k},t)\equiv \mathfrak{h}_{00}(\vec{k},t) + \mathfrak{h}^{*}_{00}(-\vec{k},t), \hspace{.5cm} \mathcal{A}_{ii}(\vec{k},t)\equiv \mathfrak{h}_{ii}(\vec{k},t) + \mathfrak{h}^{*}_{ii}(-\vec{k},t), \nonumber \\
&&\mathcal{A}_{0k}(\vec{k},t)\equiv \mathfrak{h}_{0k}(\vec{k},t) + \mathfrak{h}^{*}_{0k}(-\vec{k},t), \hspace{.5cm} \mathcal{A}_{mn}(\vec{k},t)\equiv \mathfrak{h}_{mn}(\vec{k},t) + \mathfrak{h}^{*}_{mn}(-\vec{k},t), \nonumber \\
&&\mathcal{B}_{mm}(\vec{k},t)\equiv \mathfrak{h}_{mm}(\vec{k},t) -
\mathfrak{h}^{*}_{mm}(-\vec{k},t), \hspace{.5cm}
\mathcal{C}_{mm}(\vec{k},t)\equiv
\frac{\omega}{2}\mathfrak{h}_{mm}(\vec{k},t) +
k_{m}\mathfrak{h}_{0m}(\vec{k},t).
\label{coeff}
\end{eqnarray}
\normalsize

It is easy to see that imposing the $(TT)$-gauge all the
coefficients in (\ref{densidadham}) and consequently only the first
term remains in (\ref{hamiltoniano}) leaving only the
$(TT)$-contribution.

Employing (\ref{separation}) we have that the 3 components of $\mathfrak{h}_{ij}(\vec{k},t)$ can be written as
{\setlength\arraycolsep{0.1em}
\begin{eqnarray}
\mathfrak{h}^{(TT)}_{jk}(\vec{k},t)=&\mathfrak{h}_{jk}&(\vec{k},t) - \frac{k_{m}k_{k}}{\mid \vec{k}\mid^{2}} \mathfrak{h}_{jm}(\vec{k},t) - \frac{k_{j}k_{l}}{\mid \vec{k}\mid^{2}} \mathfrak{h}_{lk}(\vec{k},t)
+ \frac{1}{2} \frac{k_{j}k_{l}k_{m}k_{k}}{\mid \vec{k}\mid^{4}} \mathfrak{h}_{lm}(\vec{k},t) \nonumber \\
 &-& \frac{1}{2} \mathfrak{h}_{mm}(\vec{k},t) \delta_{jk} + \frac{1}{2} \delta_{jk} \frac{k_{l}k_{m}}{\mid \vec{k}\mid^{2}} \mathfrak{h}_{lm}(\vec{k},t) + \frac{1}{2} \mathfrak{h}_{mm}(\vec{k},t) \frac{k_{j}k_{k}}{\mid \vec{k}\mid^{2}}
\end{eqnarray}
\begin{equation}
\mathfrak{h}^{(T)}_{jk}(\vec{k},t)=
\frac{1}{2}\left(\mathfrak{h}_{mm}(\vec{k},t) \delta_{jk} -
\delta_{jk} \frac{k_{l}k_{m}}{\mid \vec{k}\mid^{2}}
\mathfrak{h}_{lm}(\vec{k},t) - \mathfrak{h}_{mm}(\vec{k},t)
\frac{k_{j}k_{k}}{\mid \vec{k}\mid^{2}} +
\frac{k_{j}k_{l}k_{m}k_{k}}{\mid \vec{k}\mid^{4}}
\mathfrak{h}_{lm}(\vec{k},t)  \right)
\end{equation}
\begin{equation}
\mathfrak{h}^{(L)}_{jk}(\vec{k},t)= \frac{k_{m}k_{k}}{\mid \vec{k}\mid^{2}} \mathfrak{h}_{jm}(\vec{k},t) + \frac{k_{j}k_{l}}{\mid \vec{k}\mid^{2}} \mathfrak{h}_{lk}(\vec{k},t) - \frac{k_{j}k_{l}k_{m}k_{k}}{\mid \vec{k}\mid^{4}} \mathfrak{h}_{lm}(\vec{k},t).
\end{equation}
Then under an infinitesimal diffeomorphism we have
\begin{equation}
\mathfrak{h}_{ij}(\vec{k},t) \to
\mathfrak{h}'_{ij}(\vec{k},t)=\mathfrak{h}_{ij}(\vec{k},t) +
\xi_{i,j} + \xi_{j,i}.
\end{equation}
Therefore each part transforms in the following way
\begin{equation}
\mathfrak{h'}^{(TT)}_{jk}(\vec{k},t) = \mathfrak{h}^{(TT)}_{jk}(\vec{k},t) + i
\left( \delta_{jk} - \frac{k_{k}k_{j}}{\mid \vec{k}\mid^{2}}\right) k_{m}\xi_{m},
\end{equation}
\begin{equation}
\mathfrak{h'}^{(T)}_{jk}(\vec{k},t) = \mathfrak{h}^{(T)}_{jk}(\vec{k},t) - i \left( \delta_{jk} -
\frac{k_{k}k_{j}}{\mid \vec{k}\mid^{2}}\right) k_{m}\xi_{m},
\end{equation}
\begin{equation}
\mathfrak{h'}^{(L)}_{jk}(\vec{k},t) = \mathfrak{h}^{(L)}_{jk}(\vec{k},t) + \xi_{j,k} + \xi_{k,j}.
\end{equation}
Thus it is evident that the sum
\begin{equation}
\mathfrak{h'}^{(TT)}_{jk}(\vec{k},t) + \mathfrak{h'}^{(T)}_{jk}(\vec{k},t) = \mathfrak{h}^{(TT)}_{jk}(\vec{k},t) + \mathfrak{h}^{(T)}_{jk}(\vec{k},t)
\end{equation}
is invariant under these gauge transformations given by an
infinitesimal diffeomorphism. The longitudinal $(L)$-part produces
only an additional longitudinal term in $\mathfrak{h}_{ij}$ of the
form
\begin{equation}
\int d^{3}k \bigg\{ i
k_{j}(c_{k}\big(\vec{k},t)+c_{k}^{*}(-\vec{k},t)\big) + i k_{k}
(c_{j}\big(\vec{k},t)+c_{j}^{*}(-\vec{k},t)\big)\bigg\}e^{i\vec{k}\cdot\vec{x}}
.
\end{equation}
Thus the gauge transformation does not change
$\mathfrak{p}^{ij}(\vec{k},t)$, $H$ and any other $(TT)$-observable.

\vskip 1truecm
\noindent
{\it Canonical coordinates}

Similarly to other cases dealing with fields one can introduce the
canonical coordinates and their conjugate momenta
$(Q_{ij}(\vec{k}),P^{ij}(\vec{k}))$\footnote{In the literature about
linearized gravity these coordinates are better known as
$(Q^{(+)}_{ij}(\vec{k}),Q^{(-)}_{ij}(\vec{k}))$, see for instance
\cite{Kuchar:1970mu}.}. Their relation with the $\mathfrak{h}$ and
$\mathfrak{h}^*$ variables is given as usual
\begin{equation}
{Q}_{ij}(\vec{k}) := \sqrt{ \hbar \over 2 \omega(\vec{k})} \bigg(
\mathfrak{h}^*_{ij}(\vec{k}) + \mathfrak{h}_{ij}(\vec{k}) \bigg), \
\ \  {P}^{ij}(\vec{k}) := i  \sqrt{ \hbar \omega(\vec{k}) \over 2
}\bigg( \mathfrak{h}^*_{ij}(\vec{k}) - \mathfrak{h}_{ij}(\vec{k})
\bigg),
\end{equation}
or
\begin{equation}
\mathfrak{h}_{ij}(\vec{k}):= \sqrt{\omega(\vec{k}) \over 2 \hbar}
\bigg(Q_{ij}(\vec{k}) + {i \over \omega(\vec{k})}P_{ij}(\vec{k})
\bigg), \ \ \ \mathfrak{h}^*_{ij}(\vec{k}):= \sqrt{\omega(\vec{k})
\over 2 \hbar} \bigg(Q_{ij}(\vec{k}) - {i \over
\omega(\vec{k})}P_{ij}(\vec{k}) \bigg).
\end{equation}

The Poisson brackets are now
$$
\{ Q_{ij}(\vec{k},t), P^{kl}(\vec{k}',t) \}_{PB} = \delta^{ \ \
kl}_{ij} \delta(\vec{k}-\vec{k}'),
$$
\begin{equation}
\{ Q_{ij}(\vec{k},t), Q_{kl}(\vec{k}',t) \}_{PB} = 0 = \{
P^{ij}(\vec{k},t), P^{kl}(\vec{k}',t) \}_{PB}.
\end{equation}

All quantities previously discussed in preceding subsections can be
written in terms of $(Q,P)$ coordinates. For instance, the field
$h_{ij}$ is given by
\begin{equation}
h_{ij}(\vec{x},t) =  \int {d^3k \over (2 \pi)^{3/2}}
\bigg(Q_{ij}(\vec{k}) \cos\big(\vec{k}\cdot \vec{x} -
\omega(\vec{k})t \big) - {P^{ij}(\vec{k}) \over \omega(\vec{k})}
\sin \big(\vec{k} \cdot \vec{x} - \omega(\vec{k})t\big) \bigg).
\label{achecos}
\end{equation}
The momentum is given by
\begin{equation}
\pi^{ij}(\vec{x},t) = \int {d^3k \over (2 \pi)^{3/2}}  \bigg(
\omega(\vec{k})Q_{ij}(\vec{k}) \sin \big(\vec{k} \cdot \vec{x} -
\omega(\vec{k})t\big) + P_{ij}(\vec{k}) \cos\big(\vec{k}\cdot
\vec{x} - \omega(\vec{k})t \big) \bigg). \label{pecos}
\end{equation}
These coordinates also admit a decomposition in terms of the $(TT)$,
$(T)$ and $(L)$ parts
$$
Q_{ij}(\vec{k}) = Q^{(TT)}_{ij}(\vec{k}) + Q^{(T)}_{ij}(\vec{k})  +
Q^{(L)}_{ij}(\vec{k}),
$$
\begin{equation}
P_{ij}(\vec{k}) = P^{(TT)}_{ij}(\vec{k}) + P^{(T)}_{ij}(\vec{k})  +
P^{(L)}_{ij}(\vec{k}).
\end{equation}

\section{Deformation Quantization of Linearized Gravitational Field}

\noindent
{\it Deformation quantization in the temporal gauge}

We deal now with the quantization of the linearized gravitational
field in terms of the Weyl-Wigner-Groenewald-Moyal (WWGM)
deformation quantization formalism. This quantization is carried out
in the temporal gauge thought at the end of this section will be
briefly mentioned as look like the quantization in the $(TT)$-gauge.

Now following the results of \cite{Cordero:2011xa} we construct
first the Stratonovich-Weyl quantizer and the Moyal $\star -$product
corresponding to this system. In that reference it was used the WWGM
formalism in the Wheeler's superspace. In the present paper however
we prefer to use equivalently the Fourier dual, thus lets consider
the fields at the moment $t=0$ and employ
$\mathfrak{h}_{ij}(\vec{k},t)$ and $\mathfrak{p}^{ij}(\vec{k},t)$ as
the coordinates of the phase space ${\cal Z}_{LG}$.

According to the Weyl's rule if
$F=F[\mathfrak{h}_{ij},\mathfrak{p}^{ij}]$ denotes a functional on
the linearized gravitational field phase space ${\cal Z}_{LG}$ then
we can assign to the functional $F$ the following operator
$\widehat{F}$
\begin{equation}
\widehat{F} = {\cal W}
\big(F[\mathfrak{h}_{ij}(\vec{k}),\mathfrak{p}^{ij}(\vec{k})]
\big):= \int {\cal D} \big( {\mathfrak{p}^{ij}(\vec{k}) \over  2 \pi
\hbar}\big) {\cal D} {\mathfrak{h}_{ij}(\vec{k})}
F[\mathfrak{h}_{ij}(\vec{k}),\mathfrak{p}^{ij}(\vec{k})]
\widehat{\Omega}[\mathfrak{h}_{ij}(\vec{k}),\mathfrak{p}^{ij}(\vec{k})],
\label{wwgmcorr}
\end{equation}
where ${\cal W}$ is the Weyl correspondence map (or Weyl's image)
which is an isomorphism and $\widehat{\Omega}$ stands for the
operator valued distribution given by
$$
\widehat{\Omega}
[\mathfrak{h}_{ij}(\vec{k}),\mathfrak{p}^{ij}(\vec{k})] = \int {\cal
D}\big({\hbar\lambda^{ij}(\vec{k}) \over 2 \pi}\big) {\cal D}
\mu_{ij}(\vec{k}) \exp \bigg \{ -i \int d^3k
\bigg(\lambda^{ij}(\vec{k}) \mathfrak{h}_{ij}(\vec{k}) +
\mu_{ij}(\vec{k}) \mathfrak{p}^{ij}(\vec{k}) \bigg) \bigg \}
$$
\begin{equation}
\times \exp\bigg\{i\int d^3k \bigg(\lambda^{ij}(\vec{k})
\widehat{\mathfrak{h}}_{ij}(\vec{k}) + \mu_{ij}(\vec{k})
\widehat{\mathfrak{p}}^{ij}(\vec{k})\bigg)\bigg\},
\label{Stratonovich}
\end{equation}
where $\widehat{\mathfrak{h}}_{ij}(\vec{k})$ and $
\widehat{\mathfrak{p}}^{ij}(\vec{k})$ are the field operators. These
operators satisfy the commutation relations
\begin{equation}
[\widehat{\mathfrak{h}}_{ij}(\vec{k}),  \widehat{\mathfrak{p}}^{kl}(\vec{k}')] = i \hbar \delta^{ \ \ kl}_{ij} \delta(\vec{k} - \vec{k}'),
\end{equation}
and they are defined such that acting on the corresponding states
$|\mathfrak{h}_{ij}(\vec{k})\rangle$ and
$|\mathfrak{p}^{ij}(\vec{k})\rangle$ fulfils the following relations
\begin{equation}
\widehat{\mathfrak{h}}_{ij}(\vec{k})|\mathfrak{h}_{ij}(\vec{k})\rangle
= \mathfrak{h}_{ij}(\vec{k})|\mathfrak{h}_{ij}(\vec{k})\rangle,
\hspace{.5cm}
\widehat{\mathfrak{p}}^{ij}(\vec{k})|\mathfrak{p}^{ij}(\vec{k})\rangle
= \mathfrak{p}^{ij}(\vec{k})|\mathfrak{p}^{ij}(\vec{k})\rangle
\end{equation}
and the completeness property that
\begin{eqnarray}
\int {\cal D} \mathfrak{h}_{ij}(\vec{k}) |\mathfrak{h}_{ij}(\vec{k})\rangle \langle \mathfrak{h}_{ij}(\vec{k})|
= \widehat{1},  \nonumber \\
\int {\cal D} \big({\mathfrak{p}^{ij}(\vec{k}) \over 2 \pi
\hbar}\big) |\mathfrak{p}^{ij}(\vec{k})\rangle \langle
\mathfrak{p}^{ij}(\vec{k})|= \widehat{1},
\end{eqnarray}
where $\widehat{1}$ stands for the identity operator.

These operators can be separated into three parts according to Eq.
(\ref{separation})
\begin{eqnarray}
\widehat{\mathfrak{h}}_{ij}(\vec{k}) =  \widehat{\mathfrak{h}}_{ij}^{(TT)}(\vec{k}) +
 \widehat{\mathfrak{h}}_{ij}^{(T)}(\vec{k}) + \widehat{\mathfrak{h}}_{ij}^{(L)}(\vec{k}),  \nonumber \\
\widehat{\mathfrak{p}}^{ij}(\vec{k}) =
\widehat{\mathfrak{p}}^{ij(TT)}(\vec{k}) +
\widehat{\mathfrak{p}}^{ij(T)}(\vec{k}) +
\widehat{\mathfrak{p}}^{ij(L)}(\vec{k}),
\end{eqnarray}
such that their corresponding states can be written as
\begin{eqnarray}
|\mathfrak{h}_{ij}(\vec{k})\rangle = |\mathfrak{h}_{ij}^{(TT)}(\vec{k})\rangle \otimes |\mathfrak{h}_{ij}^{(T)}(\vec{k})\rangle \otimes |\mathfrak{h}_{ij}^{(L)}(\vec{k})\rangle, \nonumber  \\
|\mathfrak{p}^{ij}(\vec{k})\rangle =
|\mathfrak{p}^{ij(TT)}(\vec{k})\rangle \otimes
|\mathfrak{p}^{ij(T)}(\vec{k}) \rangle \otimes
|\mathfrak{p}^{ij(L)}(\vec{k})\rangle .
\end{eqnarray}

The commutation relations for this operators are given by
$$
[\widehat{\mathfrak{h}}_{ij}^{(TT)}(\vec{k}),
\widehat{\mathfrak{p}}^{kl(TT)}(\vec{k}')] = i\hbar \delta_{ij}^{ \
\ kl} \delta(\vec{k}-\vec{k}'),
$$
\begin{equation}
[\widehat{\mathfrak{h}}_{ij}^{(T)}(\vec{k}),
\widehat{\mathfrak{p}}^{kl(T)}(\vec{k}')] = i\hbar \delta_{ij}^{ \ \
kl} \delta(\vec{k}-\vec{k}'), \hspace{0.5cm}
[\widehat{\mathfrak{h}}_{ij}^{(L)}(\vec{k}),
\widehat{\mathfrak{p}}^{kl(L)}(\vec{k}')] = i\hbar \delta_{ij}^{ \ \
kl} \delta(\vec{k}-\vec{k}'),
\end{equation}
while all the others are equal to zero.

The operator $\widehat{\Omega}$ defined in Eq.(\ref{Stratonovich}) is the linearized gravitational field equivalent of the Stratonovich-Weyl quantizer employed in the deformation quantization of classical mechanics and we will proceed to call it in the same way.

\vskip 1truecm
\noindent
{\it The Star-Product}

We can obtain now the Moyal $\star$-product for the linearized
gravitational field. Let
$F[\mathfrak{h}_{ij}(\vec{k}),\mathfrak{p}^{ij}(\vec{k})]$ and
$G[\mathfrak{h}_{ij}(\vec{k}),\mathfrak{p}^{ij}(\vec{k})]$ be two
functionals on ${\cal Z}_{LG}$, and let $\widehat{F}$ and
$\widehat{G}$ be their corresponding operators given by the
relations
\begin{equation}
F[\mathfrak{h}_{ij}(\vec{k}),\mathfrak{p}^{ij}(\vec{k})]= {\cal
W}^{-1}(\widehat{F}), \ \ \
G[\mathfrak{h}_{ij}(\vec{k}),\mathfrak{p}^{ij}(\vec{k})]={\cal
W}^{-1}(\widehat{G}),
\end{equation}
where ${\cal W}^{-1}$ stands for the inverse Weyl correspondence.

In a similar way we can look for the functional which corresponds to
the product of the operators $\widehat{F} \widehat{G}$. We will
denote this functional by $(F \star
G)[\mathfrak{h}_{ij}(\vec{k}),\mathfrak{p}^{ij}(\vec{k})]$ and
following \cite{Cordero:2011xa} we will also call it the Moyal
$\star$-product. Then we define the star product as
\begin{equation}
\label{stardef} (F \star
G)[\mathfrak{h}_{ij}(\vec{k}),\mathfrak{p}^{ij}(\vec{k})]:= {\cal
W}^{-1}(\widehat{F} \widehat{G}) = {\rm Tr} \bigg\{
\widehat{\Omega}[\mathfrak{h}_{ij},\mathfrak{p}^{ij}] \widehat{F}
\widehat{G} \bigg\}.
\end{equation}

Substituting (\ref{wwgmcorr}) into (\ref{stardef}) and after
performing some calculations one obtain the following result

\begin{equation}
\big(F \star
G\big)[\mathfrak{h}_{ij}(\vec{k}),\mathfrak{p}^{ij}(\vec{k})] =
 F[\mathfrak{h}_{ij}(\vec{k}),\mathfrak{p}^{ij}(\vec{k})] \exp
\bigg({i\hbar\over 2} \buildrel{\leftrightarrow} \over {\cal
P}_{LG}\bigg)
G[\mathfrak{h}_{ij}(\vec{k}),\mathfrak{p}^{ij}(\vec{k})],
\end{equation}
where $\buildrel{\leftrightarrow}\over {\cal P}_{LG}$ denotes the
Poisson operator for the linearized gravity and it is given by
\begin{equation}
\buildrel{\leftrightarrow}\over {\cal P}_{LG} := \int d^3k
\bigg({{\buildrel{\leftarrow}\over {\delta}}\over \delta
\mathfrak{h}_{ij}(\vec{k})} {{\buildrel{\rightarrow}\over
{\delta}}\over \delta \mathfrak{p}^{ij}(\vec{k})} -
{{\buildrel{\leftarrow}\over {\delta}}\over \delta
\mathfrak{p}^{ij}(\vec{k})} {{\buildrel{\rightarrow}\over
{\delta}}\over \delta \mathfrak{h}_{ij}(\vec{k})}\bigg).
\end{equation}
This operator can be expressed in terms of the ($TT$), ($T$) and
($L$) components of $\mathfrak{h}_{ij}$ and $\mathfrak{p}^{ij}$ as
follows
\begin{eqnarray}
\buildrel{\leftrightarrow}\over {\cal P}_{LG} &=& \int d^3k
\bigg({{\buildrel{\leftarrow}\over {\delta}}\over \delta
\mathfrak{h}_{ij}^{(TT)}(\vec{k})} {{\buildrel{\rightarrow}\over
{\delta}}\over \delta \mathfrak{p}^{ij(TT)}(\vec{k})} -
{{\buildrel{\leftarrow}\over {\delta}}\over \delta
\mathfrak{p}^{ij(TT)}(\vec{k})}
{{\buildrel{\rightarrow}\over {\delta}}\over \delta \mathfrak{h}_{ij}^{(TT)}(\vec{k})}\bigg) \nonumber \\
&+& \int d^3k \bigg({{\buildrel{\leftarrow}\over {\delta}}\over \delta \mathfrak{h}_{ij}^{(T)}(\vec{k})} {{\buildrel{\rightarrow}\over {\delta}}\over \delta \mathfrak{p}^{ij(T)}(\vec{k})}
- {{\buildrel{\leftarrow}\over {\delta}}\over \delta \mathfrak{p}^{ij(T)}(\vec{k})}
{{\buildrel{\rightarrow}\over {\delta}}\over \delta \mathfrak{h}_{ij}^{(T)}(\vec{k})}\bigg) \nonumber \\
&+& \int d^3k \bigg({{\buildrel{\leftarrow}\over {\delta}}\over
\delta \mathfrak{h}_{ij}^{(L)}(\vec{k})}
{{\buildrel{\rightarrow}\over {\delta}}\over \delta
\mathfrak{p}^{ij(L)}(\vec{k})} - {{\buildrel{\leftarrow}\over
{\delta}}\over \delta \mathfrak{p}^{ij(L)}(\vec{k})}
{{\buildrel{\rightarrow}\over {\delta}}\over \delta
\mathfrak{h}_{ij}^{(L)}(\vec{k})}\bigg) \nonumber \\
 &=& -{i \over \hbar}\bigg\{\int d^3k
\bigg({{\buildrel{\leftarrow}\over {\delta}}\over \delta
\mathfrak{h}_{ij}^{(TT)}(\vec{k})} {{\buildrel{\rightarrow}\over
{\delta}}\over \delta \mathfrak{h}^{*ij(TT)}(\vec{k})} -
{{\buildrel{\leftarrow}\over {\delta}}\over \delta
\mathfrak{h}^{*ij(TT)}(\vec{k})}
{{\buildrel{\rightarrow}\over {\delta}}\over \delta \mathfrak{h}_{ij}^{(TT)}(\vec{k})}\bigg)  \nonumber \\
&+& \int d^3k \bigg({{\buildrel{\leftarrow}\over {\delta}}\over
\delta \mathfrak{h}_{ij}^{(T)}(\vec{k})}
{{\buildrel{\rightarrow}\over {\delta}}\over \delta
\mathfrak{h}^{*ij(T)}(\vec{k})} - {{\buildrel{\leftarrow}\over
{\delta}}\over \delta \mathfrak{h}^{*ij(T)}(\vec{k})}
{{\buildrel{\rightarrow}\over {\delta}}\over \delta \mathfrak{h}_{ij}^{(T)}(\vec{k})}\bigg) \nonumber \\
&+& \int d^3k \bigg({{\buildrel{\leftarrow}\over {\delta}}\over
\delta \mathfrak{h}_{ij}^{(L)}(\vec{k})}
{{\buildrel{\rightarrow}\over {\delta}}\over \delta
\mathfrak{h}^{*ij(L)}(\vec{k})} - {{\buildrel{\leftarrow}\over
{\delta}}\over \delta \mathfrak{h}^{*ij(L)}(\vec{k})}
{{\buildrel{\rightarrow}\over {\delta}}\over \delta
\mathfrak{h}_{ij}^{(L)}(\vec{k})}\bigg)\bigg\}. \label{moyaldos}
\end{eqnarray}
In terms of the canonical coordinates we have the Poisson operator
can be written as
\begin{eqnarray}
\buildrel{\leftrightarrow}\over {\cal P}_{LG} &=& \int d^3k
\bigg({{\buildrel{\leftarrow}\over {\delta}}\over \delta
{Q}_{ij}^{(TT)}(\vec{k})} {{\buildrel{\rightarrow}\over
{\delta}}\over \delta {P}^{ij(TT)}(\vec{k})} -
{{\buildrel{\leftarrow}\over {\delta}}\over \delta
{P}^{ij(TT)}(\vec{k})}
{{\buildrel{\rightarrow}\over {\delta}}\over \delta {Q}_{ij}^{(TT)}(\vec{k})}\bigg) \nonumber \\
&+& \int d^3k \bigg({{\buildrel{\leftarrow}\over {\delta}}\over
\delta {Q}_{ij}^{(T)}(\vec{k})} {{\buildrel{\rightarrow}\over
{\delta}}\over \delta {P}^{ij(T)}(\vec{k})} -
{{\buildrel{\leftarrow}\over {\delta}}\over \delta
{P}^{ij(T)}(\vec{k})}
{{\buildrel{\rightarrow}\over {\delta}}\over \delta {Q}_{ij}^{(T)}(\vec{k})}\bigg) \nonumber \\
&+& \int d^3k \bigg({{\buildrel{\leftarrow}\over {\delta}}\over
\delta {Q}_{ij}^{(L)}(\vec{k})} {{\buildrel{\rightarrow}\over
{\delta}}\over \delta {P}^{ij(L)}(\vec{k})} -
{{\buildrel{\leftarrow}\over {\delta}}\over \delta
{P}^{ij(L)}(\vec{k})} {{\buildrel{\rightarrow}\over {\delta}}\over
\delta {Q}_{ij}^{(L)}(\vec{k})}\bigg). \label{poissonnormal}
\end{eqnarray}

\vskip 1truecm
\noindent{\it Wigner functional}

Let denote now by $\widehat{\rho}_{phys}$ the density operator of a
physical quantum state of the linearized gravitational field, then
the Wigner functional $\rho_{_W}[\mathfrak{h}_{jk},
\mathfrak{h^{*}}_{jk}]$ corresponding to the pure state can be
obtained by
\begin{equation}
\rho_{_W}[\mathfrak{h}_{jk}(\vec{k}),
\mathfrak{h^{*}}_{jk}(\vec{k})]= {\rm Tr}
\{\widehat{\Omega}[\mathfrak{h}_{jk},
\mathfrak{h^{*}}_{jk}]\hat{\rho}_{phys}\}.
\end{equation}

In order to obtain the Wigner functional we need to solve the
constraint (\ref{gausslaw}) at the quantum level in the following
form
\begin{equation}
\nabla_{j}\widehat{\pi}^{ij}(\vec{x}) |\Psi_{phys}\rangle = 0,
\label{gausshilbert}
\end{equation}
which is a constraint on the Hilbert space of states. In similar way
as in \cite{GarciaCompean:1999bz} we introduce the equivalent form
of (\ref{gausshilbert}) in the $h$-representation of the wave
functional $\Psi_0$
\begin{equation}
\nabla_{j}{\pi}^{ij}(\vec{x}) \Psi_{0} = -i\nabla_{j}\frac{\delta\Psi_{0}}{\delta h_{ij}(\vec{x})}=0.
\end{equation}
From Eqs. (\ref{momento}) and (\ref{momentok}) we obtain that the
constraint equation yields
\begin{eqnarray}
\nabla_{j}{\pi}^{ij}(\vec{x},t)= \int {d^3k\over (2 \pi)^{3/2}}
\bigg({ \hbar \over 2 \omega(\vec{k})}\bigg)^{1/2}
\left\{\frac{-k_{j}}{2}\delta_{ij}k_{k}\big[\mathfrak{h}_{0k}(\vec{k},t)
+\mathfrak{h}^{*}_{0k}(-\vec{k},t)\big] \right. \nonumber \\
+\left.
\frac{k_{j}k_{j}}{2}\big[\mathfrak{h}_{0i}(\vec{k},t)+\mathfrak{h}^{*}_{0i}(-\vec{k},t)\big]
+
\frac{k_{j}k_{i}}{2}\big[\mathfrak{h}_{0j}(\vec{k},t)+\mathfrak{h}^{*}_{0j}(-\vec{k},t)
\big]\right\} e^{i\vec{k}\cdot\vec{x}}=0.
\end{eqnarray}
The last equation is fulfilled if and only if the following condition holds
\begin{equation}
\mathfrak{p}^{ij(L)}(\vec{k},t)+\mathfrak{p}^{*ij(L)}(-\vec{k},t)=0.
\end{equation}
Or equivalently
\begin{equation}
\mathfrak{h}^{(L)}_{0k}(\vec{k},t)+\mathfrak{h}^{*(L)}_{0k}(-\vec{k},t) =0.
\end{equation}

In this way at the quantum level the constriction equation in terms
of operators on the Hilbert space reads
\begin{equation}
[\widehat{\mathfrak{h}}_{0k}(\vec{k})+\widehat{\mathfrak{h}}^{*}_{0k}(-\vec{k})]|\Psi_{phys}\rangle
=0.
\end{equation}
Using now the deformation quantization approach the previous equation can be written as
\begin{equation}
[\mathfrak{h}_{0k}(\vec{k})+\mathfrak{h}^{*}_{0k}(-\vec{k})] \star
\rho_{W}^{(L)} =0,
\label{restriccion}
\end{equation}
which leads to
\begin{equation}
[\mathfrak{h}_{0k}(\vec{k})+\mathfrak{h}^{*}_{0k}(-\vec{k})]
\rho_{W}^{(L)} + \frac{1}{2}\left( \frac{\delta
\rho_{W}^{(L)}}{\delta\mathfrak{h}^{*}_{0k}(\vec{k})} - \frac{\delta
\rho_{W}^{(L)}}{\delta\mathfrak{h}_{0k}(-\vec{k})} \right)=0,
\end{equation}

Thus this equation will be satisfied if and only if the following
couple of equations is fulfilled
\begin{equation}
[\mathfrak{h}_{0k}(\vec{k})+\mathfrak{h}^{*}_{0k}(-\vec{k})]
\rho^{(L)}_{W} = 0
\label{unouno}
\end{equation}
and
\begin{equation}
 \frac{\delta \rho^{(L)}_{W}}{\delta\mathfrak{h}^{*}_{0k}(\vec{k})} -
\frac{\delta \rho_{W}^{(L)}}{\delta\mathfrak{h}_{0k}(-\vec{k})}  =
0.
\label{dosdos}
\end{equation}
A solution for $\rho_{W}^{(L)}$ is then given by
\begin{eqnarray}
\rho_{W}^{(L)} &\sim &  \delta(\mathfrak{p}^{ij(L)}(\vec{k})+\mathfrak{p}^{*ij(L)}(-\vec{k}))\nonumber \\
& \sim &
\delta(\mathfrak{h}_{0k}(\vec{k})+\mathfrak{h}^{*}_{0k}(-\vec{k})).
\end{eqnarray}
In terms of the $(Q,P)$-variables, the longitudinal Wigner
functional is
\begin{equation}
\rho_{W}^{(L)} \sim  \delta \big(Q_{0k}^{(L)}(\vec{k})+
Q_{0k}^{(L)}(-\vec{k})\big) \cdot \delta \big(P_{0k}^{(L)}(\vec{k})-
P_{0k}^{(L)}(-\vec{k}) \big).
\end{equation}

The ground state is defined by the following equation in the Hilbert
space
\begin{equation}
\widehat{\mathfrak{h}}^{(TT)}_{ij}(\vec{k},t)|\Psi_{phys}\rangle =
0,
\end{equation}
which in terms of the $\star$-product reads
\begin{equation}
\mathfrak{h}^{(TT)}_{ij}(\vec{k}) \star \rho^{(TT)}_{W_0} = 0.
\label{aniquilacion}
\end{equation}
With the aid of expression (\ref{moyaldos}) we get
\begin{equation}
\mathfrak{h}^{(TT)}_{ij}(\vec{k}) \rho^{(TT)}_{W_0} +
\frac{1}{2}\frac{\delta
\rho^{(TT)}_{W_{0}}}{\delta\mathfrak{h}^{(TT)*}_{ij}(\vec{k})}=0.
\end{equation}
Finally, the solution to this last equation is
\begin{equation}
\rho^{(TT)}_{W_0}[\mathfrak{h}_{ij}^{(TT)},\mathfrak{h}^{*ij(TT)}] =
A \exp\left(- 2 \int d^3k
\mathfrak{h}^{*ij(TT)}(\vec{k})\mathfrak{h}^{(TT)}_{ij}(\vec{k})\right),
\end{equation}
where $A$ stands by a normalization constant. It is easy to write
the solution in the normal coordinates $(Q,P)$ which reads
\begin{equation}
\rho^{(TT)}_{W_0} = A \exp\left\{- {1 \over \hbar} \int d^3k {1
\over \omega(\vec{k})}
\bigg(P^{(TT)}_{ij}(\vec{k})P^{ij(TT)}(\vec{k})
+\omega^2(\vec{k})Q^{(TT)}_{ij}(\vec{k})Q^{ij(TT)}(\vec{k})\bigg)\right\}.
\end{equation}

Thus the general solution $\rho_{W_0}$ simultaneously satisfying
Eqs. (\ref{restriccion}) and (\ref{aniquilacion}) can be written as
$$
\rho_{{W}_0}= \rho_{{W}_0}^{(TT)} \cdot \rho_{W}^{(L)}
$$
\begin{equation}
=  A\exp\left(- 2 \int d^3k
\mathfrak{h}^{*ij(TT)}(\vec{k})\mathfrak{h}^{(TT)}_{ij}(\vec{k})\right)\cdot
\delta\big(\mathfrak{h}^{(L)}_{0k}(\vec{k},t)+\mathfrak{h}^{*(L)}_{0k}(-\vec{k},t)
\big)
\label{wigneraches}
\end{equation}
or in terms of $(Q,P)$-variables we have
$$
\rho_{W_0}=  A \exp\left\{- {1 \over \hbar} \int d^3k {1 \over
\omega(\vec{k})} \bigg(P^{(TT)}_{ij}(\vec{k})P^{ij(TT)}(\vec{k})
+\omega^2(\vec{k})Q^{(TT)}_{ij}(\vec{k})Q^{ij(TT)}(\vec{k})\bigg)\right\}
$$
\begin{equation}
\times \delta \big(Q_{0k}^{(L)}(\vec{k})+
Q_{0k}^{(L)}(-\vec{k})\big) \cdot \delta \big(P_{0k}^{(L)}(\vec{k})-
P_{0k}^{(L)}(-\vec{k}) \big).
\label{wignernormal}
\end{equation}

\vskip 1truecm
\noindent
{\it Normal ordering}

When we are dealing with fields at the quantum level it is usual to
employ the so called normal ordering of field operators. This
ordering can be extended within the deformation quantization
formalism defining a proper operator acting over the functionals on
the phase space. For the case of the linearized gravitational field
we denote it by $\widehat{{\cal N}}^{(TT)}$ which acts on the
$(TT)$-gauge invariant functionals on the phase space ${\cal
Z}_{LG}$. It is written as

\begin{eqnarray}
\widehat{{\cal N}}^{(TT)} &:= & \exp\left\{-\frac{1}{2} \int d^3k
\frac{\delta^{2}}{\delta \mathfrak{h}^{(TT)}_{ij}(\vec{k}) \delta
\mathfrak{h}^{*ij(TT)}(\vec{k})} \right\} \nonumber \\
 & = & \exp\left\{-\frac{1}{2} \int d^3k
\frac{\delta^{2}}{\delta P^{(TT)}_{ij}(\vec{k}) \delta
P^{ij(TT)}(\vec{k})} + {1 \over \omega(\vec{k})}
\frac{\delta^{2}}{\delta Q^{(TT)}_{ij}(\vec{k}) \delta
Q^{ij(TT)}(\vec{k})} \right\}.
\end{eqnarray}

In this way, if ${\cal
O}^{(TT)}[\mathfrak{h}^{(TT)}_{ij}(\vec{k},t),\mathfrak{h}^{*ij(TT)}(\vec{k},t)]$
represents any gauge invariant functional on ${\cal Z}_{LG}$ and
$\widehat{{\cal O}}^{(TT)}$ its corresponding Weyl's image i.e. for
operator $\widehat{\cal O}^{(TT)}={\cal W}\big({\cal
O}^{(TT)}[\mathfrak{h}^{(TT)}_{ij}(\vec{k},t),\mathfrak{h}^{*ij(TT)}(\vec{k},t)]\big)$
the normal ordering operator associated is given by
\begin{eqnarray}
:\widehat{\cal O}^{(TT)}: = {\cal W}\big(\widehat{{\cal
N}}^{(TT)}{\cal
O}[\mathfrak{h}_{ij}(\vec{k},t),\mathfrak{p}^{ij}(\vec{k},t)]\big).
\end{eqnarray}

\vskip 1truecm
\noindent
{\it Correlation functions}

Let ${\cal O}$ be an invariant quantum observable and let
$\widehat{\rho}_{0 \ phys}$ be the density operator of a physical
ground state. Then one quickly finds that the expected value in the
ground state $|0\rangle$ reads
\begin{eqnarray}
\langle 0|\widehat{\cal O} |0\rangle & = & {{\rm Tr} \{
\widehat{\cal O}
\widehat{\rho}_0\}  \over {\rm Tr} \{ \widehat{\rho}_0 \} }  \nonumber \\
&= & {\int  {\cal D} ({\mathfrak{p}^{ij} \over 2 \pi \hbar}) {\cal
D}\mathfrak{h}_{ij} {\cal W}^{-1}(\widehat{\cal O})
{\rho}_{_{{W}_0}}[\mathfrak{h}_{ij},\mathfrak{p}^{ij}] \over \int
{\cal D} ({\mathfrak{p}^{ij} \over 2 \pi \hbar}) {\cal
D}\mathfrak{h}_{ij}
{\rho}_{_{{W}_0}}[\mathfrak{h}_{ij},\mathfrak{p}^{ij}]}.
\end{eqnarray}
The correlation function of $N$ observables ${\cal
O}_a=\widehat{\cal
O}[{h}_{ij}(\vec{x}_a,t_a),{\pi}^{ij}(\vec{x}_a,t_a))$, with
$a=1,\cdots , N$ can be expressed  in the deformation quantization
approach by
\begin{eqnarray}
\langle {\cal O}_1 \cdots {\cal O}_N\rangle & = & \langle
0|\widehat{\cal O}_1 \cdots \widehat{\cal O}_N |0\rangle \nonumber
\\
&=& {\int  {\cal D}( {\mathfrak{p}^{ij} \over 2 \pi \hbar}) {\cal
D}\mathfrak{h}_{ij} {\cal W}^{-1}(\widehat{\cal O}_1) \star \cdots
\star {\cal W}^{-1}(\widehat{\cal O}_N)
{\rho}_{_{{W_0}}}[\mathfrak{h}_{ij},\mathfrak{p}^{ij}] \over \int
{\cal D}\mathfrak{h}_{ij} {\cal D} ({\mathfrak{p}^{ij} \over 2 \pi
\hbar}){\rho}_{_{{W}_0}}[\mathfrak{h}_{ij},\mathfrak{p}^{ij}]}.
\end{eqnarray}
where ${\cal W}^{-1}(\widehat{\cal O}_a) ={\cal O}_a$. Of course the
same is valid in the normal coordinates $(Q,P)$ system
\begin{eqnarray}
\langle {\cal O}_1 \cdots {\cal O}_N\rangle &=& \langle
0|\widehat{\cal O}_1 \cdots \widehat{\cal O}_N |0\rangle  \nonumber \\
&=& {\int {\cal D}({P^{ij} \over 2 \pi \hbar}) {\cal D}Q_{ij}  {\cal
O}_1 \star \cdots \star {\cal O}_N {\rho}_{_{{W}_0}}[Q_{ij},P^{ij}]
\over \int {\cal D}({P^{ij} \over 2 \pi \hbar}) {\cal D}Q_{ij}
{\rho}_{_{{W}_0}}[Q_{ij},P^{ij}]}. \label{propagador}
\end{eqnarray}

\vskip .5truecm
\noindent {\it Propagator}

Once the correlation functions are defined in this formalism we
would like to compute the propagator of the graviton. In order to
perform the functional integration it is convenient to use the
scheme of canonical coordinates. It is easy to see that the
factorization of the Wigner functional (\ref{wignernormal}) and the
consideration of a gauge invariant observable\footnote{Another
important observable in the theory of gravitational waves is the
Isaacson energy tensor $T^{GW}_{\alpha \beta}= \langle
h^{(TT)}_{ij,\alpha}h^{ij(TT)}_{ \ \ ,\beta} \rangle$, which is very
important to estimate the energy flux that carries the gravitational
wave to be detected. The quadrupole expansion of the gravitational
radiation in the $(TT)$-gauge also produces important observable
quantities relevant in the detection of these waves.} in the
following form
\begin{equation}
{\cal O}^{(TT)}_{ijkl}  = h^{(TT)}_{ij} (\vec{x},t) \star
{h}^{(TT)}_{kl}(\vec{x}',t')
\label{observableuno}
\end{equation}
lead Eq. (\ref{propagador}) reduces to
\small
\begin{equation}
\langle 0| \widehat{h}^{(TT)}_{ij}(\vec{x},t) \cdot
\widehat{h}^{(TT)}_{kl}(\vec{x}',t')  |0\rangle = {\int  {\cal D}
({P^{(TT)ij}(\vec{k}) \over 2 \pi \hbar}) {\cal D}Q^{(TT)}
_{ij}(\vec{k}) h^{(TT)}_{ij} (\vec{x},t) \star
{h}^{(TT)}_{kl}(\vec{x}',t') {\rho}^{(TT)}_{_{W_0}}[ P^{ij},Q_{ij}]
\over \int {\cal D} ({P^{(TT)ij}(\vec{k}) \over 2 \pi \hbar})  {\cal
D}Q^{(TT)}_{ij}(\vec{k}) {\rho}^{(TT)}_{_{W_0}}[ P^{ij},Q_{ij}]}.
\label{goodpropa}
\end{equation}
\normalsize
Here in order to make the procedure simpler we proceed
to compute (\ref{propagador}) with only the first term in
(\ref{observableuno}).

In order to compute this correlation function one has to compute
first ${h}^{(TT)}_{ij}(\vec{x},t) \star
{h}^{(TT)}_{kl}(\vec{x}',t')$. From (\ref{achecos}) one can observe
that it is given by
$$
{h}^{(TT)}_{ij}(\vec{x},t) \star {h}^{(TT)}_{kl}(\vec{x}',t') = \int
{d^3k \over (2 \pi)^{3/2}} {d^3k' \over (2 \pi)^{3/2}}
$$
$$
\times \bigg\{Q^{(TT)}_{ij}(\vec{k}) \star Q^{(TT)}_{kl}(\vec{k}')
\cos \Theta \cos\Theta' - Q^{(TT)}_{ij}(\vec{k}) \star
{P^{(TT)}_{kl}(\vec{k}') \over \omega(\vec{k}')} \cos\Theta \sin
\Theta'
$$
\begin{equation}
- {P^{(TT)}_{ij}(\vec{k}) \over \omega(\vec{k})} \star
Q^{(TT)}_{kl}(\vec{k}')\sin \Theta \cos \Theta' +
{P^{(TT)}_{ij}(\vec{k}) \over \omega(\vec{k})} \star
{P^{(TT)}_{kl}(\vec{k}') \over \omega(\vec{k}')} \sin \Theta \sin
\Theta' \bigg\}
\label{acheproduct}
\end{equation} where $\Theta= \vec{k} \cdot \vec{x} -
\omega(\vec{k}) t$ and $\Theta'= \vec{k}' \cdot \vec{x}' -
\omega(\vec{k}') t'$. Performing the star product with the aid of
(\ref{poissonnormal}) we observe that the first and the fourth terms
have not corrections in $\hbar$, while the second and the third ones
do. From the second term we have
\begin{equation}
Q^{(TT)}_{ij}(\vec{k}) \star {P^{(TT)}_{kl}(\vec{k}') \over
\omega(\vec{k}')} = Q^{(TT)}_{ij}(\vec{k}) {P^{(TT)}_{kl}(\vec{k}')
\over \omega(\vec{k}')} +{i \hbar \over 2 \omega(\vec{k}')}
\delta_{ik} \delta_{jl} \delta(\vec{k}-\vec{k}')
\label{alfauno}
\end{equation}
and similarly for the third term
\begin{equation}
{P^{(TT)}_{ij}(\vec{k}) \over \omega(\vec{k})} \star
Q^{(TT)}_{kl}(\vec{k}')=  {P^{(TT)}_{ij}(\vec{k}) \over
\omega(\vec{k})}  Q^{(TT)}_{kl}(\vec{k}') - {i \hbar \over 2
\omega(\vec{k})} \delta_{ik} \delta_{jl} \delta(\vec{k}-\vec{k}').
\label{betados}
\end{equation}
Once we substitute (\ref{alfauno}) and (\ref{betados})  into
(\ref{acheproduct}) and subsequently into (\ref{goodpropa}), the
first term in (\ref{alfauno}) and (\ref{betados}) contributes to
(\ref{goodpropa}) as an odd integral of the form $\int dx
\exp(-ax^2) \cdot x$ and consequently the them vanish. Thus the only
contribution to the integral (\ref{goodpropa}) come from the second
terms of (\ref{alfauno}) and (\ref{betados}). In this case we have a
factor which is independent on the canonical variables $(Q,P)$ and
therefore can be factor out from the path integral giving the same
gaussian integral in the numerator and the denominator and
consequently they going to cancel out. Thus the second and third
terms give rise to a contribution to integral (\ref{goodpropa}) of
the form
\begin{equation}
 \int {d^3k \over (2 \pi)^{3}} {i\hbar \over 2 \omega(\vec{k})}
\delta_{ik} \delta_{jl} \cdot \sin\bigg(\vec{k} \cdot (\vec{x} -
\vec{x}') - \omega(\vec{k})(t-t') \bigg).
\end{equation}

The star product from the first and fourth terms from
(\ref{acheproduct}) have no correction in $\hbar$. For the first
term its is gaussian in variable $P$ and will cancel out with the
denominator. The integral in $Q$ is of the form $\int dx \exp(- a
x^2) \ x \cdot x$ and it is not zero. For the fourth term the
$Q$-integral cancel out and the integral in $P$ is also of the form
$\int d({p \over 2 \pi \hbar}) \exp(- a p^2) \ p \cdot p$ and it
gives a non-zero contribution. Therefore the total contribution from
first and fourth terms is
\begin{equation}
 \int {d^3k \over (2 \pi)^{3}} {\hbar \over 2 \omega(\vec{k})}
\delta_{ik} \delta_{jl} \cdot \cos \bigg(\vec{k} \cdot(\vec{x} -
\vec{x}') - \omega(\vec{k})(t-t') \bigg).
\end{equation}

Gathering all, one realizes that the total contribution to the
integral (\ref{goodpropa})  gives finally the propagator and it is
given by
\begin{equation}
\langle 0|  \widehat{h}^{(TT)}_{ij}(\vec{x},t) \cdot
\widehat{h}^{(TT)}_{kl}(\vec{x}',t') |0 \rangle = \int {d^3k \over
(2 \pi)^{3}}{\hbar \over 2 \omega(\vec{k})} \delta_{ik} \delta_{jl}
\exp\bigg\{ i \bigg(\vec{k} \cdot (\vec{x}-\vec{x}') -
\omega(\vec{k})(t-t') \bigg)  \bigg\}.
\end{equation}
This clearly a non-covariant propagator and it is the obtained here.
Moreover considering the symmetry of the indices $(ij)$ and $(kl)$
and the traceless condition of $h_{ij}$ we have
\begin{eqnarray}
\langle 0|  \widehat{h}^{(TT)}_{ij}(\vec{x},t) \cdot
\widehat{h}^{(TT)}_{kl}(\vec{x}',t') |0 \rangle &=& \int {d^3k \over
(2 \pi)^{3}}{\hbar \over 2 \omega(\vec{k})} \big(\delta_{ik}
\delta_{jl} + \delta_{il} \delta_{jk} -\delta_{ij} \delta_{kl}\big)
\nonumber \\
 && \times \exp\bigg\{ i \bigg(\vec{k} \cdot (\vec{x}-\vec{x}') -
\omega(\vec{k})(t-t') \bigg)  \bigg\}.
\end{eqnarray}

Moreover from this last equation the covariant propagator can be
reconstructed by taking into account the time ordering
\begin{equation}
G_{ijkl}(x,x') = \theta(t-t')\langle 0| \widehat{h}_{ij}(x) \cdot
\widehat{h}_{kl}(x') |0 \rangle + \theta(t'-t)\langle
0|\widehat{h}_{ij}(x') \cdot \widehat{h}_{kl}(x) |0 \rangle,
\end{equation}
where $x$ is the four-vector $x=(x^0,\vec{x})$, $k=(k^0,\vec{k})$
and $k \cdot x \equiv k_\mu x^\mu$. Thus we have
\begin{eqnarray}
G_{ijkl}(x,x') &=& \int {d^3k \over (2 \pi)^3}{\hbar \over 2
\omega(\vec{k})} \big(\delta_{ik} \delta_{jl} + \delta_{il}
\delta_{jk} -\delta_{ij} \delta_{kl} \big) \nonumber \\
&& \times \bigg[ \theta(t-t')\exp\big\{ ik \cdot (x-x')\big\} +
\theta(t'-t)\exp\big\{ -ik \cdot (x-x') \big\} \bigg],
\end{eqnarray}
where $\theta(t)$ is the Heaviside function. It is an easy matter to
see that the covariant propagator reads
\begin{equation}
G_{\mu \nu \rho \sigma}(x,x') = \hbar \int {d^4 k \over (2\pi)^4}
\bigg({\eta_{\mu \rho} \eta_{\nu \sigma} + \eta_{\mu \sigma}
\eta_{\nu \rho} - \eta_{\mu \nu} \eta_{\rho \sigma} \over k^2}
\bigg){\exp\big\{ ik \cdot (x-x')\big\}}.
\end{equation}
This is precisely the graviton propagator in the literature found by
the deformation quantization method.

\vskip 1truecm
\noindent {\it Deformation quantization in the $(TT)$-gauge}

Now we briefly discuss the deformation quantization of the
linearized gravity in the $(TT)$-gauge (\ref{normaTT}) or
(\ref{gausslaw}). In this case the Poisson brackets are given by
(\ref{poissonTT}) in terms of the traceless transverse delta
function (\ref{deltaTT}). Moreover $h_{ij}(\vec{x},t)$ and
$\pi^{ij}(\vec{x},t)$ are no longer independent variables. Thus the
suitable variables are
$(\mathfrak{h}^{(TT)}_{ij},\mathfrak{h}^{*ij(TT)})$ or
$({Q}^{(TT)}_{ij},{P}^{ij(TT)})$. Furthermore the Gauss law
constraint (\ref{gausslaw}) is automatically satisfied. Consequently
it is eliminated the longitudinal parts from all formulas obtained
along the paper. Then the resulting equations constitute the
formulas for the deformation quantization of the linearized gravity
in the $(TT)$-gauge.

\vskip 1truecm
\section{Final Remarks}
In this paper we have applied the WWGM formalism to the linearized
gravitational field. The Weyl's correspondence is explicitly
developed and the Stratonovich-Weyl quantizer, and the star product
are constructed. These results were used to obtain the Wigner
funtional $\rho_{W_0}$ for the ground state. It is found that this
functional factorizes into two parts (see Eqs. (\ref{wigneraches})
and (\ref{wignernormal})) one is the $(TT)$-component
$\rho^{(TT)}_{W_{0}}$ and other one $\rho^{(L)}_{W_{0}}$ depending
only on the longitudinal part $(L)$. Thus within the deformation
quantization formalism we recover the same form for the Wigner
functional of the ground state found by Kuchar \cite{Kuchar:1970mu}
in the context of canonical quantization and by Hartle
\cite{Hartle:1984ke} in the path integral quantization.

Since deformation quantization is the quantum mechanics on the phase
space our formulation was explicitly non-covariant, however
similarly to the electromagnetic case in the temporal gauge, the
formulation is shown to be covariant at the end of the quantization
procedure.

In addition the normal ordering and the correlation functions were
defined and the explicit computation of the propagator for the
graviton was explicitly performed in Sec. 3.

Deformation quantization is a well established procedure in quantum
mechanics, however in the context of field theory it still must be
confirmed thought working out examples of theories of different
nature. It is the goal of the present paper to continue exploring
such possibilities in order to test this formalism and bring more
evidence of the validity of the formalism to work in the future more
complicated cases. We consider that deformation quantization
possesses various advantages in order to deal with these complicated
problems as to treat systems with phase spaces topologically
non-trivial or with curved phase spaces. For these cases the
canonical quantization could lead to the existence of non-hermitian
operators which is avoided in deformation quantization as a result
of the use of classical objects instead of operators. For these
reasons more examples and further research is needed to develop this
approximation.

\vskip 2truecm
\centerline{\bf Acknowledgments}

The work of H. G.-C. and F. J. T. was partially supported by
SNI-M\'exico, CONACyT research grants: 103478 and 128761. In
addition F. J. T. was partially supported by COFAA-IPN and by
SIP-IPN grant 20110968.

\vskip 1truecm

\bibliography{octaviostrings}
\addcontentsline{toc}{section}{Bibliography}
\bibliographystyle{TitleAndArxiv}


\end{document}